\documentclass[aps,twocolumn,showpacs,amsmath,amssymb]{revtex4}
\usepackage{natbib}
\begin{document}
\title{Comment on "Self-dual teleparallel formulation of general
relativity and the positive energy theorem" G. Y. Chee}
\author{V. Pelykh}
\email{pelykh@lms.lviv.ua} \affiliation{Pidstryhach Institute Applied Problems in
Mechanics and Mathematics Ukrainian National Academy of Sciences, 3B
Naukova Str.,\\
Lviv, 79601, Ukraine}
\date{February 21, 2005}
\begin{abstract}
We give a correct tensor proof of the positive energy problem for the case
including momentum on basis of conditions of existence of the  two-to-one
correspondence between the Sen-Witten spinor field and the Sen-Witten
orthonormal frame. These conditions were obtained in our previous
publications, but true significance of our works was not estimated
properly by G.Y.Chee, and these were not correct quoted in his
publication. On other hand, the main result of our work is key
argument in favour of geometrical nature of the Sen-Witten spinor field.
\end{abstract}
 \maketitle

 Ever since Witten developed the spinor method to
prove the
  positive energy theorem (PET) for gravity
 the problem of comparing this method
   with the tensor methods has been a subject of
continuing interest.  Goldberg's initial
    categorical negation of the possibility that connections exist
between these two
    methods  \cite{goldberg} "For the first time spinors have an
intrinsic role for which tetrads
    cannot be substituted" was partially disproved  by Dimakis
    and M\"uller-Hoissen \cite{dim1}, and later by Frauendiener
\cite{frau}.
     Dimakis and M\"uller-Hoissen supposed that the spinor field
     could be "replaced" by some orthonormal frame field, so that the
existence of a
     global solution  to the Sen--Witten equation
  would imply the existence of globally
     defined orthonormal frames on the Cauchy surface.  But in general
the solution
     to the Sen--Witten equation will have zeros; from this
     Dimakis and M\"uller-Hoissen concluded that each orthonormal
frame field, as well as
    Nester's special orthonormal frame field (SOF, triad) on a spacelike
hypersurface in
     an asymptotically Minkowskian manifold   can exist almost everywhere only
     \cite{dim1,dim2}.

     Frauendiener established that a correspondence may exist   between
     the spinor field $\lambda^A$, which satisfies on a spacelike
hypersurface $\Sigma$
     the Sen-Witten equation (SWE)
     \[\mathcal{D}^A{}_B\lambda^B=0, \]
and a triad, which satisfies on $\Sigma$ a certain gauge condition, and noted that this
gauge is closely related to Nester's.
     But this Frauendiener result is valid only under the additional
assumption that the Sen-Witten spinor field has no zeros.

     Nester's SOF  consists of the variables  that describe the
physical degrees of freedom  in general relativity. Analogously, the preferred lapse
     $N=\lambda_A\lambda^{+A}\equiv \lambda$ and shift
     $N^a=-\sqrt2\mathrm{i}\lambda^{+(A}\lambda^{B)}$,
     constructed by Ashtekar and Horowitz \cite{ash} from the Witten
spinor, give an especially simple
     form of gravitational Hamiltonian. Nevertheless, degeneracy  of
Nester's SOF or
     Ashtekar and Horowitz preferred time variables, which is  due to the
     existence of zeros of the spinor field, and may occur
     on subsets of dimensions lower that 3  on the Cauchy hypersurface,
puts the physical sense of  these two constructions in doubt.
      Taking this degeneracy into account, Nester \cite{nesCQG} had
supposed
      that a SOF exists at least for geometries in a neighborhood of
Euclidean space.
     Chee in his paper \cite{chee} states that the Nester gauge
condition can  be derived from
      Witten's equation without any additional conditions for all
geometries,
      even on non-maximal hypersurfaces.
     Below we prove that this statement is not valid without
additional assumptions
     and give a corrected proof of  the PET for the case including momentum.

     Indeed, the
correspondence between the spinor field, which satisfies the Sen-Witten equation, and a
triad, which satisfies  a certain gauge condition, is correctly defined by the Sommers
transformation \cite{som}
\begin{equation}\label{1}
\theta^1=\frac{\sqrt2}{2\lambda}(L+\overline L),\quad \theta^2=\frac{\sqrt2}{2\lambda
i}(L-\overline L), \quad\theta^3=\widetilde L,
  \end{equation}
where $\theta^a$ is a coframe basis, $ L=-\lambda_A\lambda_B,\quad
\lambda=\lambda_A\lambda^{A+}$, and $\quad \widetilde L=\mid
L\mid^{-1}*\left(L\land\overline L\right)$
if and only if the spinor field $\lambda_A$
vanishes nowhere on $\Sigma$.
     This follows from the fact that the bilinear form
\[\frac1{\sqrt2}n^{A{\dot A}}\lambda_A\overline{\lambda}_{\dot
A}=\lambda_A\lambda^{A+}\equiv \lambda,\] where $n$ is the unit normal one-form to
$\Sigma$, is Hermitian positive definite, and $\lambda$ does not vanish at a point on
$\Sigma$ if  the solution
  $\lambda_A$ does not have a zero at this point. But $\lambda_A$ is
the solution of the SWE, which is of elliptic type; zeros of
solutions  to such equations not only  may, but must exist, and
these have a  clear physical meaning: for example, zeros of
solution to the equation for vibrations of a flat membrane
  are the node lines of standing waves.

In Chee's work the possible existence of node manifolds for the SWE is not excluded  but
it is ignored completely --- there even is no mention of the assumption
$\chi^2\equiv\lambda\equiv\lambda_A\lambda^{A+}\neq0$.
  As a result, the Sommers transformation (1), written by Chee as formula
(47), does not exist on node manifolds, and, consequently  his  conditions (48), which
are Nester's conditions, are not fulfilled. Then equation (51) for the boundary term
\[\oint_S \widetilde{B}^{(AB)}dS_{AB}\]
is not fulfilled, the choice $N=\lambda_A\lambda^{A+}$ is not possible, and this means
that the last formula (53) of publication \cite{chee}
\begin{eqnarray*}
\oint_S \widetilde{B}^{(AB)}dS_{AB}=2\int_\Sigma \sigma\left(\nabla
^{(BC)}\lambda^{A}\right)^+\left(\nabla_{(BC)}\lambda_A\right)dV
\\+4\pi G\int_\Sigma \sigma\lambda^{+A}\left(T_{00}\lambda_A+\sqrt 2
T_{0AB}\lambda^B\right)dV
\end{eqnarray*}
in general is not correct
  \footnote{For an absolutely correct proof it is
also necessary to  abandon   the application of  a
three-dimensional truncation of the four-dimensional Gauss theorem
\cite{nesterPhysLet}.}.


   We now give
the corrected proof of  the PET  for the case including momentum
on the basis of conditions  for the existence of  the
correspondence between Nester's gauge and the SWE, obtained by us
in publications \\ \cite{pelykhJMP,pelykhJPA,pelykhCQG}.

  {\bf Definition 1.} {\it A point where the solution for an
elliptic system of equations is equal to zero is called  a node point
   of the solution.  }

  From the general theory of elliptic differential equations it is
known that nontrivial solutions cannot vanish on an open subdomain, but they can become
zero on subsets of lower dimensions $k,\, k=0,1,...n-1$, where $n$ is the dimension of
the domain.

{\bf Definition 2.} {\it A node submanifold of dimension $s, s=1,2,...n-1$,
  is a maximal connected subset of dimension $s$
   consisting of  node points of the solution.}

 In
the case of a single selfadjoint elliptic equation in $V^3$
  the node submanifolds can only be surfaces  that divide the
domain, but in the case of a system of equations the topology of
node submanifolds has greater variety: it can be also that of
lines or of points.

  Let us consider first the case when the Cauchy hypersurface is
maximal.

  {\bf Theorem 1.} \cite{pelykhJMP} {\it Let $\lambda^C$ satisfy
Reula's condition \cite{reula} and be a solution of the SWE with an asymptotically flat
initial data set, satisfying the dominant energy condition. Then on a maximal
hypersurface $\Sigma$, the solution $\lambda^C$ is everywhere free from node point .}

   On the
basis of this theorem we obtain

{\bf Theorem 2.} {\it  Let an initial data set $(h_{\mu\nu}, {\cal K}_{\pi\rho})$  on a
maximal hypersurface $\Sigma$ be asymptotically flat and satisfy the dominant energy
condition. Then everywhere on $\Sigma$ the Sen--Witten equation with Reula conditions for
the spinor field \cite{reula} and Nester's gauge are equivalent (up to sign of the
spinor).}

  Therefore, in this simple case of maximal hypersurface the Chee
proof is correct,
  if the hypersurface  is asymptotically flat and the
dominant energy condition  is fulfilled.

   To investigate the node manifolds of the SWE
on non-maximal hypersurfaces in \cite {pelykhJPA}
  we had developed an oscillation theory  for general
double--covariant systems \textbf{\footnote{We call a system of
equations double-covariant if it is covariant under  arbitrary
transformations of coordinates in $V^3$, and covariant under local
$SU(2)$ transformations  in a local space that is isomorphic to
the complexified tangent space in every point to $V^3$.}}
 of
elliptic equations of 2nd order in $\mathbb{R}^3$. Applying it in the same work to the
SWE   for solutions of the form $\lambda^C=\lambda^C_{\infty}+\beta^C$, where
$\lambda^C_{\infty}$ is an asymptotically covariant constant spinor field on $\Sigma$,
and $\beta^C$ is an element of the Hilbert space
  ${\cal H}$, defined in \cite{reula} (these conditions for solution
we  call
  the Reula conditions for the spinor field), we had obtained the
following theorem:

{\bf Theorem 3.} {\it Let:

a) the initial data set be asymptotically flat;

b) the matrix of the spinorial tensor
\[
C_A{}^B:=\frac{\sqrt2}{4}D_{A}{}^{B}{\cal K}+\frac14\varepsilon _A{}^B\left({\cal
K}^2+\frac12{\cal K}_{\pi\rho}{\cal K}^{\pi\rho}+\mu\right)
\]
   have everywhere on $\Sigma$ at least one non-negative eigenvalue,
for definiteness $C_0$;

c) ${\rm Re}\,\lambda^0_{\infty}$ or ${\rm Im}\,\lambda^0_{\infty}$ be asymptotically
nowhere equal to zero.

Then the asymptotically constant nontrivial solution $\lambda^C$ of the SWE does not have
  node points on $\Sigma$.}

This theorem  allowed us to prove  in Theorem 4 the existence everywhere on $\Sigma$ of a
certain class of orthonormal three-frames, which generalize Nester's special three-frame
(Sen--Witten orthonormal three-frame, SWOF). This class of SWOF satisfies the gauge
conditions
\begin{eqnarray}
\varepsilon^{abc}\omega_{abc}\equiv*q=0,\qquad\omega^a{}_{1a}\equiv-\widetilde q
_1=F_1,\nonumber\\\label{2}
  \omega^a{}_{2a}=-\widetilde q_2=F_2, \qquad\omega^a{}_{3a}
=-\widetilde q_3={\cal K}+F_3,
\end{eqnarray}
where $\omega_{abc}$ are the connection one-form coefficients, and
  $F=\mathrm{d}\ln\lambda$ ; conditions
(\ref{2}) coincide with Nester's gauge if and only if the one-form ${\cal
K}\lambda^{+(A}\lambda^{B)}$ is exact.

  {\bf Theorem 4.}  {\it Let  conditions of Theorem 3 be fulfilled.
Then
    a two-to-one correspondence between the Sen-Witten spinor and the
Sen-Witten orthonormal frame exists everywhere on $\Sigma$.}

  That is why in the case of non-maximal hypersurfaces  the  tensor
proof of the PET for
   the case including momentum is valid
only if conditions a) and b) of our Theorem 3  are fulfilled.

{\bf Acknowledgments.} I wish to thank D. Brill for a helpful discussion.

\bibliography{pelykh6}

\end{document}